\input harvmac
\def\ra{\rightarrow}
\def\epsK{\varepsilon_K}
\def\Hefft{{\cal H}_{\rm eff}^{\Delta b=2}}
\def\Heffo{{\cal O}^{\Delta b=1}}
\def\MVIA{M_{12}^{\rm VIA}}

\def\O{{\cal O}}
\def\tBK{\tilde B_K}
\def\oB{\omega_B}
\def\oK{\omega_K}
\def\of{\omega_f}
\def\Re{{\rm Re}}
\def\Im{{\rm Im}}

\def\YGTitle#1#2{
\nopagenumbers\abstractfont\hsize=\hstitle\rightline{#1}%
\vskip .6in\centerline{\titlefont #2}\abstractfont\vskip .3in\pageno=0}
\YGTitle{\vbox{
   \hbox{SLAC-PUB-7623}
   \hbox{WIS-97/20/Aug-PH}
   \hbox{NSF-PT-97-3}
   \hbox{hep-ph/9708398}}}
{\vbox{\centerline{The Role of the Vacuum Insertion Approximation}
\vbox{\vskip 0.truecm}
\vbox{\centerline{in Calculating CP Asymmetries in $B$ Decays}}}}
\smallskip
\centerline{Yuval Grossman,$^a$ Boris Kayser$\,^b$ and Yosef Nir$\,^c$}
\smallskip
\bigskip
\centerline{\it $^a$Stanford Linear Accelerator Center,
 Stanford University, Stanford, CA 94309, USA}
\centerline{\it $^b$National Science Foundation,
 4201 Wilson Boulevard, Arlington, VA 22230, USA}
\centerline{\it $^c$Department of Particle Physics,
 Weizmann Institute of Science, Rehovot 76100, Israel}
\smallskip
\bigskip
\baselineskip 18pt
\noindent

CP asymmetries in $B$ decays into final CP eigenstates are
in many cases theoretically clean. In particular, they do not depend
on the values of hadronic parameters. The sign of the asymmetries,
however, does depend on the sign of the $B_B$ parameter. Furthermore,
the information from $\epsK$ that all angles of the unitarity triangles
lie in the range $\{0,\pi\}$ depends on the sign of the $B_K$ parameter.
Consequently, in the (unlikely) case that the vacuum insertion
approximation is such a poor approximation that either $B_B$ or $B_K$
is negative, the sign of CP asymmetries in neutral $B$ decays will
be opposite to the standard predictions. Various subtleties
concerning the role of $K-\bar K$ mixing in the case of final states
with a single $K_S$ or $K_L$, such as the $B\ra\psi K_S$
decay, are clarified.
\bigskip

\baselineskip 18pt
\leftskip=0cm\rightskip=0cm

\Date{}


\newsec{Introduction and Formalism}

CP asymmetries in $B$ decays into final CP eigenstates
\nref\CaSa{A.B. Carter and A.I. Sanda, Phys. Rev. Lett. 45 (1980) 952;
 Phys. Rev. D23 (1981) 1567.}%
\nref\BiSa{I.I. Bigi and A.I. Sanda, Nucl. Phys. B193 (1981) 85;
 B281 (1987) 41.}%
\nref\DuRo{I. Dunietz and J.L. Rosner, Phys. Rev. D34 (1986) 1404.}%
\refs{\CaSa-\DuRo}\ will provide stringent tests of the Kobayashi-Maskawa
mechanism of CP violation. For decay processes that depend on a single
CKM phase, such as the $B\ra\psi K_S$ mode, the Standard Model
prediction is theoretically very clean (for reviews, see {\it e.g.}
\nref\NiQurev{Y. Nir and H.R. Quinn,
 Ann. Rev. Nucl. Part. Sci. 42 (1992) 211.}%
\nref\Kays{B. Kayser,
 in Trieste HEP Cosmology 1995, p.432 (hep-ph/9702264).}%
\nref\BuFl{A.J. Buras and R. Fleischer, hep-ph/9704376.}%
\refs{\NiQurev-\BuFl}). In particular,
while the magnitude of neutral meson mixing amplitudes, namely
$\Delta m_B$ and $\Delta m_K$, suffers from large hadronic
uncertainties in the matrix elements (parameterized, respectively,
by $B_Bf_B^2$ and $B_K$), the CP asymmetries are independent of the
value of these parameters. It is a little known fact, however, that the
{\it sign} of the asymmetries does depend on the sign of $B_B$ and,
in an indirect way
\ref\NiQu{Y. Nir and H.R. Quinn, Phys. Rev. D42 (1990) 1473.},
also on the sign of $B_K$. In this work we explain
how this dependence arises and describe the consequences in
(the unlikely) case that the vacuum insertion approximation is
surprisingly poor so that it gives the wrong sign of the matrix elements.

Before we start a detailed and technical analysis of the sign
dependence of the otherwise clean CP asymmetries, we give the
general argument for the existence of this dependence.
In the decays of neutral $B$ mesons to CP eigenstates, the
CP violating asymmetry arises solely from an interference between an
amplitude which involves $B-\bar B$ mixing, and one which does not. The
relative phase of these two interfering amplitudes includes the sign of
the hadronic matrix element for $B-\bar B$ mixing. Since this matrix
element is determined by the CP conserving strong interactions, its sign
is the same in the decay of a $B^0_{\rm phys}(t)$ and in that of a
$\bar B^0_{\rm phys}(t)$. A reversal of this sign would obviously
reverse the sign of the contribution of the interference term to both
the decay rate for $B^0_{\rm phys}(t)$ and the decay rate for
$\bar B^0_{\rm phys}(t)$. Thus, a reversal of the sign of the hadronic
matrix element would cause a reversal of the CP violating asymmetry
between these two decay rates.

As there are many subtle points in this discussion, we repeat here
the analysis of CP violation in $B$ and $K$ decays with particular
attention to signs. We focus on the neutral $B$ meson system, but
the analysis in this section applies equally well to the neutral
$K$ system. Our phase convention is defined by
\eqn\phacon{{\rm CP}|B^0\rangle=\oB|\bar B^0\rangle,\ \ \
{\rm CP}|\bar B^0\rangle=\oB^*|B^0\rangle,\ \ \ (|\oB|=1).}
Physical observables do not depend on the phase factor $\oB$.
We define $q$ and $p$ to be the components of the neutral $B$
interaction eigenstates in the mass eigenstates,
\eqn\defqp{|B_{1,2}\rangle\ =\ p|B^0\rangle\pm q|\bar B^0\rangle.}
We further define
\eqn\defMGot{M_{12}-{i\over2}\Gamma_{12}\equiv\vev{B^0|\Hefft|\bar B^0},}
where $M$ and $\Gamma$ are hermitian matrices, so that
\eqn\hermit{M_{12}^*=M_{21},\ \ \ \Gamma_{12}^*=\Gamma_{21}.}
The mass and width difference between the physical states are given by
\eqn\defDelMG{\Delta m\equiv M_2-M_1,\ \ \ \Delta\Gamma\equiv
\Gamma_2-\Gamma_1.}
Solving the eigenvalue equations gives
\eqn\solveMG{\eqalign{
(\Delta m)^2-{1\over4}(\Delta\Gamma)^2=&(4|M_{12}|^2-|\Gamma_{12}|^2),\cr
\Delta m\Delta\Gamma=&4\Re(M_{12}\Gamma_{12}^*),\cr}}
\eqn\solveqp{{q\over p}=-{2M_{12}^*-i\Gamma_{12}^*\over
\Delta m-{i\over2}\Delta\Gamma}.}
The quantity $(q/p)$ plays an important role in the calculation
of CP asymmetries in neutral $B$ decays and will introduce, as
we shall see, some dependence on hadronic physics.

\newsec{The Vacuum Insertion Approximation}

The effective Hamiltonian that is relevant to $M_{12}$ is of the form
\eqn\Htwo{\Hefft\propto e^{+2i\phi_B}[\bar d \gamma^\mu(1-\gamma_5) b]^2+
e^{-2i\phi_B}[\bar b\gamma^\mu(1-\gamma_5) d]^2}
where $2\phi_B$ is a CP violating (weak) phase. (We use the Standard
Model $V-A$ amplitude, but the results can be generalized to any Dirac
structure.) For example, within the Standard Model
\eqn\phiBSM{\phi_B=\arg(V_{tb}V_{td}^*).}
(We implicitly assume here that long distance contributions to
$B-\bar B$ mixing are negligible.)

The $M_{12}$ matrix element is often calculated in the vacuum
insertion approximation (VIA):
\eqn\VIA{\MVIA=\vev{B^0|\Heffo|0}\vev{0|\Heffo|\bar B^0},}
where
\eqn\Hone{\Heffo\propto e^{+i\phi_B}[\bar d \gamma^\mu(1-\gamma_5) b]+
e^{-i\phi_B}[\bar b \gamma^\mu(1-\gamma_5) d].}
Under CP transformations,
\eqn\CPLL{\bar\psi_i\gamma^\mu(1-\gamma_5)\psi_j\ \ra\
-\bar\psi_j\gamma_\mu(1-\gamma_5)\psi_i,}
thus we learn that
\eqn\VIAone{\vev{0|\Heffo|\bar B^0}=- \oB^* e^{2i\phi_B}
\vev{0|\Heffo|B^0}.}
From the hermiticity of $\Heffo$ we know that $\vev{B^0|\Heffo|0}=
\vev{0|\Heffo|B^0}^*$. This fact, in combination with \VIA\ and
\VIAone, gives
\eqn\VIAf{\MVIA=- \oB^* e^{2i\phi_B}|\MVIA|.}
The ratio between the true value of $M_{12}$ and its value in the VIA
is conventionally parameterized by a factor $B_B$:
\eqn\BsubB{M_{12}=- \oB^* e^{2i\phi_B}B_B|\MVIA|.}
As the strong interactions conserve CP, the $B_B$ parameter is real.
Yet its sign could a-priori be positive or negative.

\newsec{The CP Asymmetries in $B\ra D^+D^-$ and $B\ra\psi K_S$}

To see how the various phases and signs affect calculations of
CP violation, we consider CP asymmetries in neutral $B$ decays
into final CP eigenstates:
\eqn\defaCP{
a_{f_{CP}}={\Gamma(B^0_{\rm phys}(t)\ra f_{CP})-
\Gamma(\bar B^0_{\rm phys}(t)\ra f_{CP})\over
\Gamma(B^0_{\rm phys}(t)\ra f_{CP})+
\Gamma(\bar B^0_{\rm phys}(t)\ra f_{CP})}.}
We now introduce the various ingredients that enter the calculation
of such asymmetries.

For the neutral $B$ system, we define
\eqn\defBme{\Delta m_B>0\ \ (\Longrightarrow\
|B_1\rangle\equiv|B_L\rangle,\ \ \ |B_2\rangle\equiv|B_H\rangle),}
($L$($H$) stand for light (heavy)).
Taking into account that $\Delta m_B\gg|\Delta\Gamma_B|$, eqs.
\solveMG\ and \solveqp\ simplify into
\eqn\solveMGB{\Delta m=2|M_{12}|,\ \ \
\Delta\Gamma=2\Re(M_{12}\Gamma_{12}^*)/|M_{12}|,}
\eqn\solveqpB{{q\over p}=-{M_{12}^*\over|M_{12}|}.}
Note that $q/p$ (and therefore also $a_{f_{CP}}$) is independent of
$\Delta\Gamma$. In particular, the relative sign between
$\Delta m$ and $\Delta\Gamma$ does not play a role here.
Putting \BsubB\ in \solveqpB\ we finally get
\eqn\qpB{{q\over p}= \oB e^{-2i\phi_B}{\rm sign}(B_B).}

Additional phase dependence of CP asymmetries comes from decay
amplitudes. We define $A_f$ and $\bar A_f$ according to
\eqn\defAf{A_f=\vev{f|{\cal H}_d|B^0},\ \ \
\bar A_f=\vev{f|{\cal H}_d|\bar B^0}.}
The decay Hamiltonian is of the form
\eqn\Hdecay{{\cal H}_d\propto e^{+i\phi_f}[\bar q\gamma^\mu(1-\gamma_5)d]
[\bar b\gamma_\mu(1-\gamma_5)q]+e^{-i\phi_f}
[\bar q\gamma^\mu(1-\gamma_5)b][\bar d\gamma_\mu(1-\gamma_5)q],}
where $\phi_f$ is the appropriate weak phase. (For simplicity we use a
$V-A$ decay amplitude, but the results hold for any Dirac  structure.)
From \CPLL\ we learn that under a CP transformation the two terms in
\Hdecay\ are interchanged except for the $e^{+i\phi_f}$ and
$e^{-i\phi_f}$ phase factors. Then
\eqn\AAdecay{\bar A_f=\of\oB^* e^{-2i\phi_f}A_f,}
where CP$|f\rangle=\of|\bar f\rangle$. For a final CP eigenstate,
$f=f_{CP}$, the phase factor $\of$ is replaced by $\eta_{f_{CP}}=\pm1$,
the CP eigenvalue of the final state. Then
\eqn\CPdecay{{\bar A_{f_{CP}}\over A_{f_{CP}}}=\eta_f
\oB^* e^{-2i\phi_f}.}

An important role in CP violation is played by a complex quantity
$\lambda_f$, defined by
\eqn\lambdaf{\lambda_f={q\over p}{\bar A_f\over A_f}.}
For $B$ decays into final CP eigenstates, we find from \qpB\ and
\CPdecay:
\eqn\CPlam{\lambda_{f_{CP}}=\eta_{f_{CP}}
e^{-2i(\phi_B+\phi_f)}{\rm sign}(B_B),}
which is independent of phase conventions.
The asymmetry $a_{f_{CP}}$ of eq. \defaCP\ takes a particularly simple
form when the decay amplitude is dominated by a single weak phase
\eqn\cleanaCP{a_{f_{CP}}=-\Im\lambda_{f_{CP}}\sin(\Delta m_B\,t).}
From eq. \CPlam\ we find then that
\eqn\ImCPlam{\Im\lambda_{f_{CP}}=-\eta_{f_{CP}}{\rm sign}(B_B)
\sin[2(\phi_B+\phi_f)].}

To take an example, we now calculate the CP asymmetry in $B\ra D^+D^-$.
Within the Standard Model and neglecting penguin diagrams,
the decay phase defined in \Hdecay\ is given by
\eqn\phiDSM{\phi_{D^+D^-}=\arg(V_{cd}V_{cb}^*).}
(Unlike \phiBSM, which is sensitive to new physics, for tree level
processes such as $b\ra c\bar cd$, the Standard Model tree level
diagram is likely to dominate even in the presence of new physics.
Therefore \phiDSM\ is likely to hold almost model independently.)
Using \phiBSM\ and \phiDSM, and taking into account that
$\eta_{D^+D^-}=+1$, we find for $\lambda$ defined in \CPlam:
\eqn\lamDD{\lambda_{D^+D^-}={\rm sign}(B_B)
\left({V_{tb}^*V_{td}\over V_{tb}V_{td}^*}\right)
\left({V_{cd}^*V_{cb}\over V_{cd}V_{cb}^*}\right),}
\eqn\CPDpDm{\Im\lambda_{D^+D^-}=-\sin(2\beta){\rm sign}(B_B),}
where
\eqn\defbeta{
\beta\equiv\arg\left[-{V_{cd}V_{cb}^*\over V_{td}V_{tb}^*}\right].}
Eq. \CPDpDm\ is often displayed in the literature without its dependence
on $B_B$. The reason is that it is widely believed that the vacuum
insertion approximation gives a reasonable approximation to the true
values of the relevant matrix elements. (Lattice calculations strongly
support this notion
\ref\Flyn{For a recent review, see J.M. Flynn, hep-lat/9611016.}.)
In particular, it is believed that it gives the correct sign of the
matrix elements. One should not forget, however, that
the dependence on the hadronic physics does exist.

The situation is somewhat more complicated in decays with a single
$K_S$ (or $K_L$) in the final state. There is some confusion
in the literature concerning such decays which we would like to clarify.
The three main points concerning this mode are the following:
\item{a.} In $B\ra\psi K_S$, the kaon will be experimentally identified
by its decay to two pions within roughly one $K_S$ lifetime.
\item{b.} The smallness of $\epsK$ implies that the contribution
from $K_L\ra\pi\pi$ within roughly one $K_S$ lifetime is negligible.
\item{c.} The smallness of $\epsK$ also implies that $K_S$ is almost
purely a CP-even state.

This situation allows a straightforward derivation of the asymmetry.
In particular, it implies that the relative phase between the
direct $K^0\ra\pi\pi$ amplitude and the $K^0\ra\bar K^0\ra\pi\pi$
amplitude is very small and practically does not affect the
CP asymmetry. Using the notation $\psi (2\pi)_K$ to describe
the final state, the amplitude ratio is given by
\eqn\Apsipipi{{\bar A_{\psi(2\pi)_K}\over A_{\psi(2\pi)_K}}=
\eta_{\psi(2\pi)_K}\oB^* e^{-2i\phi_{\psi(2\pi)_K}},}
where $\eta_{\psi(2\pi)_K}=-1$. The relevant phase is found simply
from the decay chain $B^0\ra\psi K^0\ra\psi(2\pi)_K$.
Within the Standard Model, but practically model-independently,
it is given by
\eqn\phipiSM{\phi_{\psi(2\pi)_K}=\arg(V_{cs}V_{cb}^*V_{us}^*V_{ud}).}
Using \Apsipipi\ and \phipiSM, we get the Standard
Model value for $\lambda_{\psi(2\pi)_K}$:
\eqn\lamPpp{\lambda_{\psi(2\pi)_K}=-{\rm sign}(B_B)
\left({V_{tb}^*V_{td}\over V_{tb}V_{td}^*}\right)
\left({V_{cs}^*V_{cb}\over V_{cs}V_{cb}^*}\right)
\left({V_{ud}^*V_{us}\over V_{ud}V_{us}^*}\right).}
Taking into account that unitarity of the three-generation
CKM matrix implies that, to a very high accuracy,
$\left({V_{ud}^*V_{us}\over V_{ud}V_{us}^*}\right)=
\left({V_{cd}^*V_{cs}\over V_{cd}V_{cs}^*}\right)$, the Standard Model
prediction for the CP asymmetry in $B\ra\psi(2\pi)_K$ is
\eqn\CPpsiKp{\Im\lambda_{\psi(2\pi)_K}=\sin(2\beta){\rm sign}(B_B).}
Notice that, to get \CPpsiKp, it is not essential whether the typical
kaon mixing time is shorter or longer than the decay time. The only
important information about $K-\bar K$ mixing is that, to an excellent
approximation, its amplitude is aligned with that of the $K\ra\pi\pi$
decay amplitude.


Another point of interest is the fact that one can learn about
new physics in $K-\bar K$ mixing from a comparison of $a_{D^+D^-}$
and $a_{\psi K_S}$
\ref\NiSi{Y. Nir  and D. Silverman, Nucl. Phys. B345 (1990) 301.}.
(We assume here that the tree contribution is dominant among the
Standard Model contributions to $B\ra D^+D^-$.)
This may seem puzzling in view of our discussion above,
where we argued that \lamPpp\ is independent
of the physics that is responsible for $K-\bar K$ mixing.
Indeed, allowing new physics in $B-\bar B$ mixing and in $K-\bar K$
mixing but not in the relevant decay processes, $b\ra c\bar cd$,
$b\ra c\bar cs$ and $s\ra u\bar ud$, we have
\eqn\DDNP{\lambda_{D^+D^-}=\left({q\over p}\right)_B\oB^*
\left({V_{cd}^*V_{cb}\over V_{cd}V_{cb}^*}\right),}
and
\eqn\PKNP{\lambda_{\psi K_S}=-\left({q\over p}\right)_B\oB^*
\left({V_{cs}^*V_{cb}\over V_{cs}V_{cb}^*}\right)
\left({V_{ud}^*V_{us}\over V_{ud}V_{us}^*}\right).}
Then, if experiments find
\eqn\findNP{a_{D^+D^-}\neq-a_{\psi K_S},}
this will necessarily require a violation of the Standard Model relation
\eqn\NPimply{\left({V_{cs}^*V_{cd}\over V_{us}^*V_{ud}}\right)\approx-1.}
However, \NPimply\ holds if either of the following two conditions is
valid:
\item{a.} The three generation CKM matrix is unitary;
\item{b.} $K-\bar K$ mixing is dominated by the Standard Model box
diagrams with intermediate charm and up quarks.

Therefore, \findNP\ will signal that (a) the quark sector
is larger than just the three standard generations, and (b) there is
a new physics contribution to $K-\bar K$ mixing.

\newsec{The Role of $K-\bar K$ Mixing}

In contrast to $B-\bar B$ mixing, long distance contributions
are potentially significant in $K-\bar K$ mixing. As we do not
know how to calculate these contributions reliably, we will just
parameterize them by
\eqn\defD{\tBK\equiv B_K\ {[M_{12}({\rm LD})+M_{12}({\rm SD})]
\over M_{12}({\rm SD})},}
where LD (SD) stand for long (short) distance. Here,
$B_K$ is the $K$-system short distance mixing parameter,
analogous to $B_B$ (see \BsubB):
\eqn\BsubK{M_{12}({\rm SD})=- \oK^* e^{2i\phi_K}B_K|\MVIA({\rm SD})|,}
where $\oK$ is defined through
\eqn\phaconK{{\rm CP}|K^0\rangle=\oK|\bar K^0\rangle,\ \ \
{\rm CP}|\bar K^0\rangle=\oK^*|K^0\rangle,\ \ \ (|\oK|=1).}
The points that we would like to emphasize,
concerning these parameters, are the following:
\item{(i)} As we saw in the last section, neither sign($\tBK$)
nor sign($B_K$) affect the sign of the CP asymmetry in $B\ra\psi K_S$,
which depends only on the fact that $K_S$ is (approximately) CP even.
\item{(ii)} Sign($\tBK$) does affect the sign of $\Delta m_K\equiv
m(K_L)-m(K_S)$, and from the positivity, experimentally,
of $\Delta m_K$, we know that sign($\tBK$)=+1.
\item{(iii)} Sign($B_K$), which is not known from experiment and need
not agree with sign($\tBK$), does play an indirect, but essential, role
in predicting the signs of CP asymmetries in neutral $B$ decays. For,
a reversal of sign($B_K$) would reverse the signs of such quantities as
$\sin2\beta$.

We now explain points (ii) and (iii) in some detail. First, we show
that the experimental fact that the heavier kaon mass eigenstate is,
to an excellent approximation, CP odd (or, equivalently, does not decay
to final two pions), namely that (ignoring CP violation)
\eqn\KLCPodd{\lambda_{K\ra\pi\pi}\equiv\left({q\over p}\right)_K
{\bar A^K_{\pi\pi}\over A^K_{\pi\pi}}=1,}
fixes the sign of $\tBK$ to be positive. Here, $q_K$ and $p_K$
are defined by
\eqn\defqpK{|K_{S,L}\rangle\ =\ p_K|K^0\rangle\pm q_K|\bar K^0\rangle,}
where $L$($S$) stand for long (short), and we have chosen
$\Delta\Gamma_K<0$.
It is experimentally known that the long-lived kaon is heavier
\ref\Hill{D.G. Hill {\it et al.},
 Phys. Rev. D4 (1971) 7, and references therein.},
namely $\Delta m_K>0$. Neglecting the CP violating effects, which are of
$\O(10^{-3})$, and going through the same analysis as in the $B$ system,
we find
\eqn\qpK{\left({q\over p}\right)_K=\oK e^{-2i\phi_K}{\rm sign}(\tBK).}
For the amplitude ratio, we have
\eqn\Kamp{{\bar A^K_{\pi\pi}\over A^K_{\pi\pi}}=
\eta_{\pi\pi}\oK^* e^{-2i\phi_{\pi\pi}}.}
where $\eta_{\pi\pi}=+1$. We get
\eqn\KLCPval{\lambda_{K\ra\pi\pi}=e^{-2i(\phi_K+\phi_{\pi\pi})}
{\rm sign}(\tBK).}

Within the Standard Model, $(M_{12})_K$ is described by box
diagrams with intermediate charm and up quarks, leading to
\eqn\phiKSM{\phi_K=\arg(V_{cs}V_{cd}^*).}
The $s\ra u\bar ud$ decay is dominated by the $W$-mediated tree diagram
(this holds model independently), leading to
\eqn\phipipi{\phi_{\pi\pi}=\arg(V_{us}^*V_{ud}).}
With three quark generations,
$\arg(V_{cs}V_{cd}^*)=\arg(V_{us}V_{ud}^*)[{\rm mod}\ \pi]$ to
within a few milliradians. (Were this not the case, we would not know
$\phi_K$ since the long-distance part involves $V_{us}V_{ud}^*$
while the dominant box diagram in the short-distance part depends
on $V_{cs}V_{cd}^*$.) Then,
\eqn\KLCPSM{\lambda_{K\ra\pi\pi}={\rm sign}(\tBK)\ \Longrightarrow\
{\rm sign}(\tBK)=+1.}

Next, we would like to ask whether we can tell the sign of $\sin2\beta$
from the existing measurements of CP violation in $K$ decays? Note that
all angles of the unitarity triangle are either in the range $\{0,\pi\}$
or in the range $\{\pi,2\pi\}$. Then, if we know sign($\sin\phi$), where
$\phi$ is any of the three angles of the unitarity triangle, then we know
sign($\sin\beta$). Furthermore, as $|V_{ub}/V_{cb}|\leq0.10$
(it suffices here that $|V_{ub}/V_{cb}|\leq\sin\theta_C=0.22$, namely
that $\beta$ is either in the range $\{0,\pi/2\}$ or $\{3\pi/2,2\pi\}$),
we learn that sign($\sin\beta$)=sign($\sin2\beta$). The question is
then whether the measurement of $\epsK$ tells us unambiguously
sign($\sin\phi$).

To find the answer, we have to analyze precisely those
$\O(10^{-3})$ effects that we neglected in \KLCPodd\ and write instead
\eqn\KLCP{\lambda_{K\ra\pi\pi}=1-2\epsK.}
(This expression holds to zeroth order in $A_2/A_0$, where $A_I$
is the decay amplitude into two pions in isospin $I$ state.
To first order in $A_2/A_0$, it is $\lambda_0=(q/p)_K(\bar A_0/A_0)$
which appears on the left hand side of \KLCP. However, this distinction
is irrelevant to our discussion here.) Naively, using
\eqn\lamPP{\lambda_{K\ra\pi\pi}={\rm sign}(\tBK)
\left({V_{cs}^*V_{cd}\over V_{cs}V_{cd}^*}\right)
\left({V_{ud}^*V_{us}\over V_{ud}V_{us}^*}\right),}
we would conclude that we get a clean determination of one small phase.
However, as $\epsK$ is of $\O(10^{-3})$,
we need to include other effects of this order that we neglected in
$M_{12}^K$, particularly the small phase difference between
$M_{12}$ and $\Gamma_{12}$ and the contributions proportional to
$V_{ts}V_{td}^*$ and $(V_{ts}V_{td}^*)^2$. After a lengthy but well-known
and straightforward calculation \BuFl, the resulting constraint is
\eqn\epsCON{{\rm sign}(B_K)\sin\gamma>0,}
where
\eqn\defgamma{
\gamma\equiv\arg\left[-{V_{ud}V_{ub}^*\over V_{cd}V_{cb}^*}\right].}

Note that it is indeed $B_K$ which appears in \epsCON\ and not
the $\tBK$ parameter defined in \defD. (The long distance
contributions to $M_{12}$ are in phase with $\Gamma_{12}$ and
therefore do not contribute.) Consequently, we cannot
say that the sign of $B_K$ is experimentally determined.
Only if the LD contribution is smaller than the SD one,
or if it is large but has the same sign as the SD one, then
sign($\tBK$)=sign($B_K$). However, if we are not willing to state that
$|M_{12}^K({\rm LD})|<|M_{12}^K({\rm SD})|$, then the Standard Model
result that $\sin\gamma>0$ depends on the validity of the
VIA at least to the extent that $B_K>0$ \NiQu.
(Lattice calculations
\nref\Shar{S. Sharpe,
 Nucl. Phys. Proc. Suppl. 53 (1997) 181 (hep-lat/9609029).}%
\refs{\Shar,\Flyn}, the $1/N$ approach
\nref\BBG{W.A. Bardeen, A.J. Buras and J.-M. G\'erard,
 Phys. Lett. B211 (1988) 343.}%
\nref\PiRa{A. Pich and E. de Rafael, Nucl. Phys. B358 (1991) 311.}%
\refs{\BBG-\PiRa}, QCD sum rules
\nref\Deck{R. Decker, Nucl. Phys. B277 (1986) 661.}%
\nref\Nari{S. Narison, Phys. Lett. B351 (1995) 369.}%
\refs{\Deck-\Nari}, and various other methods
\nref\Pide{A. Pich and E. de Rafael, Phys. Lett. B158 (1985) 477.}%
\nref\Prad{J. Prades {\it et al.}, Z. Phys. C51 (1991) 287.}%
\nref\Brun{C. Bruno, Phys. Lett. B320 (1994) 135.}%
\nref\BiPr{J. Bijens and J. Prades, Phys. Lett. B342 (1995) 331;
 Nucl. Phys. B444 (1995) 523.}%
\nref\ABFL{V. Antonelli, S. Bertolini, M. Fabbrichesi and E.I. Lashin,
 hep-ph/9610230.}%
\refs{\Pide-\ABFL}\ support $B_K>0$.)

\newsec{Conclusions}

To summarize our main points:

1. The Standard Model predictions for the values
of CP asymmetries in $B^0$ decays into final CP eigenstates
are independent of the values of hadronic parameters. However,
the sign of all asymmetries depend on the sign of $B_B$, that
is the ratio between the short distance contributions to
$B-\bar B$ mixing and their value in the vacuum insertion approximation.

2. In decays into final states with a single neutral kaon, where the
kaon is identified by its decay to two pions, there is no dependence
on the phase of $K-\bar K$ mixing. Perhaps a better way of making
this statement is to say that the relevant phase is known experimentally.

3. Still, the Standard Model predictions for the sign of the asymmetries
depends on information from $\epsK$ which does depend on the sign of
$B_K$ (the analog of $B_B$ for the $K$ system).

4. The sign of $B_K$ is not known experimentally. The experimental fact
that the heavier neutral kaon is, to an excellent approximation, CP odd,
fixes the sign of another parameter, $\tBK$, which (unlike $B_K$)
depends also on the long distance contributions to $K-\bar K$ mixing.
If long distance contributions are larger than the short distance ones,
the sign of $B_K$ could, in principle, differ from the sign of $\tBK$.

We emphasize that, while we gave the two explicit examples of $B\ra
D^+D^-$ and $B\ra\psi K_S$, the same analysis holds for any $B$ decays
into final CP eigenstates that are dominated by a single weak phase.

Very likely, the vacuum insertion approximation is a reasonable
approximation for the matrix elements of the $\Delta b=2$ and
$\Delta s=2$ four-quark operators.
However, one has to bear in mind that the Standard Model
predictions are not entirely independent of this approximation:
\item{(i)} If $B_B<0$ and $B_K>0$, all the asymmetries will have
an opposite sign to the standard prediction;
\item{(ii)} If $B_K<0$ (which requires that the long distance
contributions to $\Delta m_K$ are larger in magnitude and opposite
in sign to the short distance ones) and $B_B>0$ then, again, all the
asymmetries will have an opposite sign to the standard prediction;
\item{(iii)} If $B_B<0$ and $B_K<0$, all the asymmetries will have
the predicted sign because the two sign errors cancel.

If, as expected, experiments find $\Im\lambda_{D^+D^-}<0$
and $\Im\lambda_{\psi K_S}>0$, it will give an experimental support
(though not a completely rigorous evidence) that the vacuum
insertion approximation is a reasonable method to estimate
the matrix elements of the relevant four quark operators.

\vskip 1 cm
\centerline{\bf Acknowledgments}
We thank Isi Dunietz for useful discussions.
B.K. would like to thank Guido Altarelli for the excellent hospitality
of CERN, and Leo Stodolsky for that of the Max Planck Institut.
Y.G. is supported by the Department of Energy under contract
DE-AC03-76SF00515. Y.N. is supported in part by the United States --
Israel Binational Science Foundation (BSF), by the Israel Science
Foundation, and by the Minerva Foundation (Munich).

\listrefs
\end